\documentclass[12pt]{article}
\usepackage[dvips]{graphicx}

\def\half{{\textstyle \frac{1}{2}}}%

\newcommand{\beq}{\begin{equation}}%
\newcommand{\eeq}{\end{equation}}%
\newcommand{\beqs}{\begin{eqnarray}}%
\newcommand{\eeqs}{\end{eqnarray}}%

\def\m#1{$#1$}

\def\beq{\begin{equation}}
\def\eeq{\end{equation}}

\def\m#1{$#1$}
\def\half{{1\over 2}}
\usepackage{amssymb}
\def\beqs{\begin{eqnarray}}
\def\eeqs{\end{eqnarray}}
\def\[{\left[}
\def\]{\right]}
\def\({\left(}
\def\){\right)}

\def\pmb#1{{\mathbf #1}}
\def\text#1{#1}
\def\textit#1{{\it #1}}

\begin{document}
%\begin{frontmatter}

\centerline{\bf Exact Solution of the  Landau-Lifshitz Equations}
\centerline{\bf   for a Radiating Charged Particle  in  The  Coulomb Potential  }

\centerline{\sf  S. G. Rajeev\footnote{rajeev@pas.rochester.edu} }
\begin{center}{Department of Physics and Astronomy\\
        University of Rochester, Rochester, New York 14627\\
	}
\end{center}
\begin{abstract}
We solve exactly the classical non-relativistic Landau-Lifshitz equations of motion for a charged particle moving in  a Coulomb potential, including radiation damping. The general solution involves the Painlev\`e transcendent of type II. It confirms our physical intuition    that a negatively charged classical particle will spiral into the nucleus, supporting the the validity of the Landau-Lifshitz equation.

\end{abstract}

\newpage

\section{ Introduction}
A corner stone of theoretical physics is  the exact solution of the  motion of a particle moving in an inverse square law force. The orbits are conic sections: ellipses or hyperbolae depending on initial conditions. The original application was to the motion of planets around the Sun\cite{Newton} . Later the same problem was found to arise in the Rutherford scattering of alpha particles and in the classical model of the atom. 

The orbits of charged particles  in a Coulomb potential cannot be   conics exactly, as  it is  a fundamental principle of electrodynamics that all charged particles must radiate when accelerated\cite{Jackson,Rohrlich}.  The radiation  carries away energy, and therefore acts as a dissipative force, changing the  equation of motion. Can we still find an exact solution for the motion of the particle in a Coulomb field, after taking account of radiation reaction?

Deriving the  correct equation of motion for a charged particle, including this   radiation damping, is not a simple matter. The problem is that radiation reaction  is the force exerted on the particle by its own electromagnetic field; a straightforward application of the Lorentz force law will give an infinite force in the case of a point particle. Dirac
\cite{Dirac}  found a way through this minefield of divergences and deduced an equation of motion including the radiation reaction. A key point was that the divergences are removed by a renormalization of the mass of the charged particle.

This Lorentz-Dirac equation of motion is  a third order ODE, as the radiation reaction force is proprortional to the derivative of acceleration. Typical initial conditions will give unphysical  solutions that `runaway': the energy  grows without bound instead of decaying.  Thus Dirac's work, although a major step forward, cannot be the final word on   the equation of motion of charged particles.

 Spohn \cite{ Rohrlich,Spohn}   showed that these unphysical runaway solutions   can be avoided if the initial data lie in  a `critical manifold'; i.e., if the initial conditions are fine-tuned to avoid the runaway unphysical solutions. This turns out to be the   same  as treating  the force due to the radiation as a first order correction.  We get this way  a second order equation with physically sensible solutions. Although without  the modern  understanding, this  equation of motion for a radiating particle were  given first in the classic text of Landau and Lifshitz\cite{LandauLifshitz}. Therefore these are known now as Landau-Lifshitz (LL)  equations of motion.  See Ref. \cite{RohrlichAJP} for a physical argument in support of the LL equations.
 
 There is no unanimity yet that these are the exact equations of motion of a radiating charged particle\cite{LLSkeptic} . In addition to experimental tests, we need to verify their  theoretical consistency. As an example, it should  not be possible for a negatively charged particle to orbit a nucleus in an elliptical orbit: it should plunge into the nucleus as the radiation it emits carries away  energy and angular momentum. It is important to verify the physical correctness of the LL equations by checking that its solutions have this property.  
 
 In this paper, {\em we  solve exactly  the non-relativistic Landau-Lifshitz equations in  a  Coulomb field }; the general solution is in terms of \cite{Ince, Fokas}  Painleve transcendents of type II. The same differential equation , with different initial conditions,   appears in several other physical problems, such as the Tracy-Widom law for random matrix eigenvalues\cite{TracyWidom,  Deift}. 
 Our solution turns out to have the correct asymptotic properties: the orbit of a negatively charged particle does spiral in towards the nucleus. 

The Lorentz-Dirac equation of motion has been studied in the Coulomb field. It has a complicated, unphysical behavior, and no general exact solution is known. For example, in an  {\em attractive}  Coulomb potential there are solutions that accelerate away to infinity. See Ref. \cite{Rohrlich} Section 6-15, \cite{HuschiltBaylis} The   approximate, numerical  treatment of the radiative reaction in Ref. \cite{AguiarBarone} is closer to our physical results .

Due to quantum effects, the LL equations cannot be the right description at the short distances characteristic of atoms. Our solution  might still be an approximate description of an electron with a  large principal quantum number in an atom. It should also describe an alpha particle scattered by a nucleus and  an electron (or positron) emitted by a nucleus, all of whose motion is  affected by the radiation emitted.    Relativistic corrections will become important as the velocity of the particle grows; we are currently investigating  the exact solvability of the relativistic LL equations in a Coulomb field.

The LL equations have already been solved in a constant electric field and in a constant magnetic field\cite{Herrera}. We hope that more physically realistic situations will open up to study using the techniques we describe here. In another paper \cite{DissipativeQuantum} we have proposed a canonical formulation and  a quantum wave equation for dissipative systems of  a particular type. The case we study here happens to be of this type, so we hope that a quantum treatment of a radiating electron in an atom along these lines  is also possible. This might allow us to go beyond perturbation theory in the calculation of line-widths of a hydrogenic atom.

The analogous problem in General Relativity of a star being captured by a blackhole, its energy and angular momentum being carried away by gravitational radiation is of great importance in connection with the LIGO project to detect gravitational waves. We hope that our solution of the much simpler electrodynamic problem will help in understanding this case as well.

\section{The Landau-Lifshitz Equations}

The LL equation of motion of a radiating charged particle in an  electrostatic  field  is \cite{LandauLifshitz}, 
\beq
{d\over dt} \left[ \gamma{\mathbf v}\right]={\mathbf a} +\tau\left[\gamma( {\mathbf v}\cdot\nabla) {\mathbf a}+{{\mathbf v}\cdot {\mathbf a}\over c^2}{\mathbf a}
-{\gamma^2\over c^2} {\mathbf v}\left\{ {\mathbf a}^2-{({\mathbf v}\cdot {\mathbf a})^2\over c^2}
\right\}
\right]
\eeq
where
\beq
{\mathbf a}= {q\over m} {\mathbf E},\quad \gamma={1\over \sqrt{1-{v^2\over c^2}}}
\eeq
Also,
\beq
\tau= \frac{2}{3}\frac{q^2}{\text{mc}^3},
\eeq
 $q$ and $m$ being  the charge and mass of the particle respectively. The dissipation parameter $\tau$ has units of time; for the electron  it would be the  (2/3 )  of the classical electron radius divided by $c$.

In the non-relativistic limit,   it  is much simpler:
\beq
{d\over dt}\left[{\mathbf v}+\tau\nabla U
\right]+\nabla U=0,\quad  {\mathbf v}={d{\mathbf r}\over dt}
\eeq
where $U$ is ${q\over m}$  times  the electrostatic potential.
For a central potential,
\beq
\nabla  U=\hat{\mathbf r}U_r, \quad U_r={dU\over dr}
\eeq
\beq
\frac{d}{dt}\left[\pmb{v}+\tau  \hat{\pmb{r}}U_r\right]+ \hat{\pmb{r}}U_r=0.
\eeq
Taking the cross product with the position vector  gives, with 
$
\pmb{L}=\pmb{r}\times \pmb{v},
$
\beq
{d\pmb{L}\over dt}=-{\tau}{U_r\over r}\pmb{L}.
\eeq

Thus the  direction of angular momentum is preserved. If  the initial conditions are such that $\pmb{L}\neq 0$, the motion will lie in the plane normal to this vector. If $\pmb{L}=0$ initially, it remains zero and the motion is along a straight-line passing through the center of the potential.

Using the standard identities
\beq
\pmb{v}^2=v_r^2+ \frac{L^2}{r^2}
\eeq
\beq
v_r=\frac{dr}{dt}=\hat{r}\pmb{.}\pmb{v},\quad
 \frac{d}{dt}\hat{\mathbf r}=\frac{1}{r}\left[\pmb{v}-v_r\hat{\mathbf r}\right],\quad 
 \hat{\mathbf r} .\frac{d}{dt}\hat{\mathbf r}=0
\eeq
\beq
\frac{dv_r}{dt}=\frac{L^2}{r^3}+\hat{\pmb{r}}.\frac{d\pmb{v}}{\text{dt}}
\eeq
we get the  system of ODE 
\beq
v_r={dr\over dt},\quad
\frac{d}{\text{dt}}\left[v_r+\tau U_r\right]=\frac{L^2}{r^3} -U_r 
,\quad
{dL\over dt}=-\tau {U_r\over r}L.
\eeq
\section{The Coulomb Potential}

For the Coulomb potential $U=\frac{k}{r}, U_r=-\frac{k}{r^2}$ and 
\beq
\frac{dL}{dt}=\frac{ k\tau L}{r^3}
\eeq
Thus the centrifugal force is a total time derivative:
\beq
\frac{L^2}{r^3}=\frac{1}{2 k\tau}\frac{d}{dt}L^2
\eeq
This coincidence allows to write the radial force equation as 
\beq
{d\over dt}\left[v_r-\frac{\tau }{r^2}- \frac{1}{2k\tau }L^2\right]={k\over r^2}
\eeq
Put 
\beq
z=v_r-\frac{k\tau} {r^2}- \frac{1}{2 k \tau}L^2
\eeq
to write this as 
\beq
{dL\over dt}={k\tau \over r^3}L, \quad
{dz\over dt}={k\over r^2},\quad 
{dr\over dt}={L^2\over 2 k\tau}+z+{k\tau\over r^2}
\eeq
Using the fact that this is an autonomous system ( i.e.,  $t$ does not appear explicitly) we can eliminate $dt$,  to get a system of two ODEs,
\beq
{dL\over dz}={\tau\over r}L,\quad
{dr\over dz}=  {1\over 2k^2\tau^2}  r^2L^2+{1\over k} r^2 z+\tau
\eeq
We note as an aside that in  the case of purely radial motion, $L=0$ this reduces to a Riccati equation for $\rho={1\over r}$:
\beq
{d\rho\over dz}=-\left[{z\over k}+\tau\rho^2\right]
\eeq
This can solved in terms of Airy functions.

Returning to the general case, we can rewrite the above system as a single second order ODE:
\beq
{d^2 L\over dz^2}=-{1\over 2 k^2}L^3- {\tau\over k}z L
\eeq
Up to scaling, this is  the particular case with $\alpha=0$  of the Painleve II equation ( See Ref. \cite{Ince},   page 345 )
\beq
{d^2 u\over dx^2}=2 u^3+x u +\alpha
\eeq
Defining constants
\beq
b=\left[ -{\tau\over k}\right]^{1\over 3},   a=\sqrt{-2 k^2}\ b.
\eeq
we have the solution
\beq
L=  a u(b z).
\eeq
When $k<0$,  as for an attractive Coulomb potential, $u$ is purely imaginary and  the independent variable $x=bz$ is real. $u$ also  depends on a complex `modular'  parameter $s$ that is determined by the initial conditions \cite{Fokas}.

The identities
\beq
L=r^2 {d\theta\over dt}, \quad {dz\over dt}={k\over r^2}
\eeq
allow us to determine the polar angle:
\beq
{d\theta\over dz}={1\over k}L. 
\eeq
By a quadrature and a differentiation of  the Painleve transcendent,  $r$ and $\theta$ are both found  as functions of the  parameter $z$, determining the orbit :
\beq
r(z)=\tau {u(z)\over {du(z)\over dz}},\quad \theta(z)=\theta_1+{a\over k}\int_{z_1} ^z u(bz)dz.
\eeq

\section{ Examples}
\subsection{A Decaying  Orbit}
To plot orbits, another form of the equations is sometimes more   convenient. Define 
$y=L^2$ and change to $\theta$ as  the independent variable:
\beq
{dL\over d\theta}={k\tau\over r},\quad {d {1\over r}\over d\theta}=-{L\over 2k\tau}-{z\over L}-{k\tau\over Lr^2}
\eeq
\beq
{dL^2\over d\theta}=2k\tau {L\over r},\quad {d\over d\theta}\left[{L\over r}\right]=-{L^2\over 2k\tau}-z
\eeq
to get the third order ODE:
\beq
{d^3y\over d\theta^3} +{dy\over d\theta} +{2k^2\tau \over \surd y}=0.\label{yODE}
\eeq
The orbit is then given by 
\beq
{1\over r(\theta)}={1\over k\tau}{d\surd y\over d\theta}
\eeq
 We can find the orbit   by numerically integrating the above third order ODE. This was done before it was realized that the equation can be solved analytically.  It is still useful as a way to   plot  a slowly decaying  orbit:

   \includegraphics{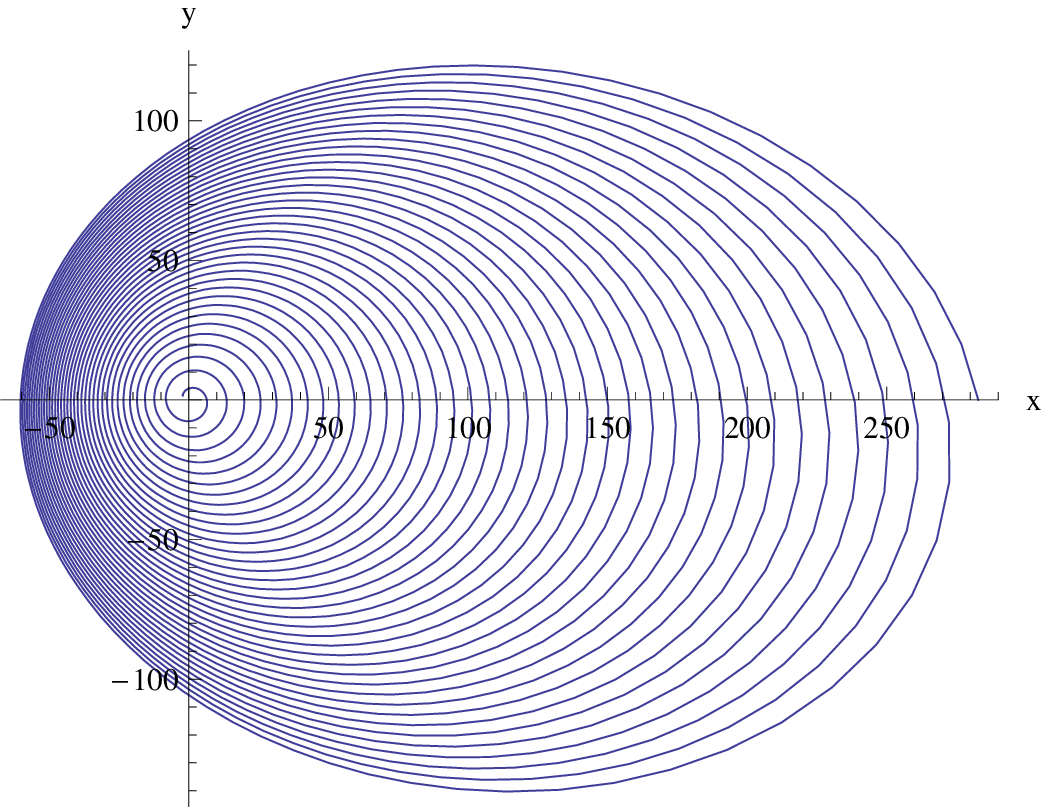}

\subsection{A Capture Orbit}
 We want a solution\footnote{
For simplicity, we use in this section   the  natural units  of the problem, with $|k|=\tau=1$. Normal units can be restored by dimensional analysis.} for which $ r={u\over u'}>0$  .  Note that 
${dz\over dt}<0$.  If $r(z)$ vanishes it must be a simple zero, since near a zero   ${dr\over dz}\sim 1$.  Thus $r\sim z-z_0$,  $L\sim C (z-z_0)$ where  $C$ and $z_0$  are real  constants of integration.

 If a particle approaches from infinity with angular momentum $L_1$ and radial velocity  $v_1$, 
 \beq
   z_1=v_1+ \half L_1^2.
  \eeq
   Since ${dr\over dz}\sim v_1 r^2$, we get 
 \beq
 r\sim {1\over v_1(z-z_1)}
 \eeq
 Thus \m{z_1} is a simple zero of \m{u'(z)}.
 
 In other words,  to describe the capture of an incoming charged particle by an attractive Coulomb potential, we just have to solve the Painleve II equation  with the initial conditions
 \beq
z_1= v_1+\half L_1^2,\quad  u(z_1)={1\over \sqrt{-2}} L_1,\quad u'(z_1)=0
 \eeq
 and evolve to a point $z_0<z_1$ at which $u(z_0)=0$. The complete orbit corresponds to the finite range $z_0\leq z\leq z_1$ in the parameter $z$. We plot an example, obtained by numerical calculations,  of such a capture orbit  below.

   \includegraphics{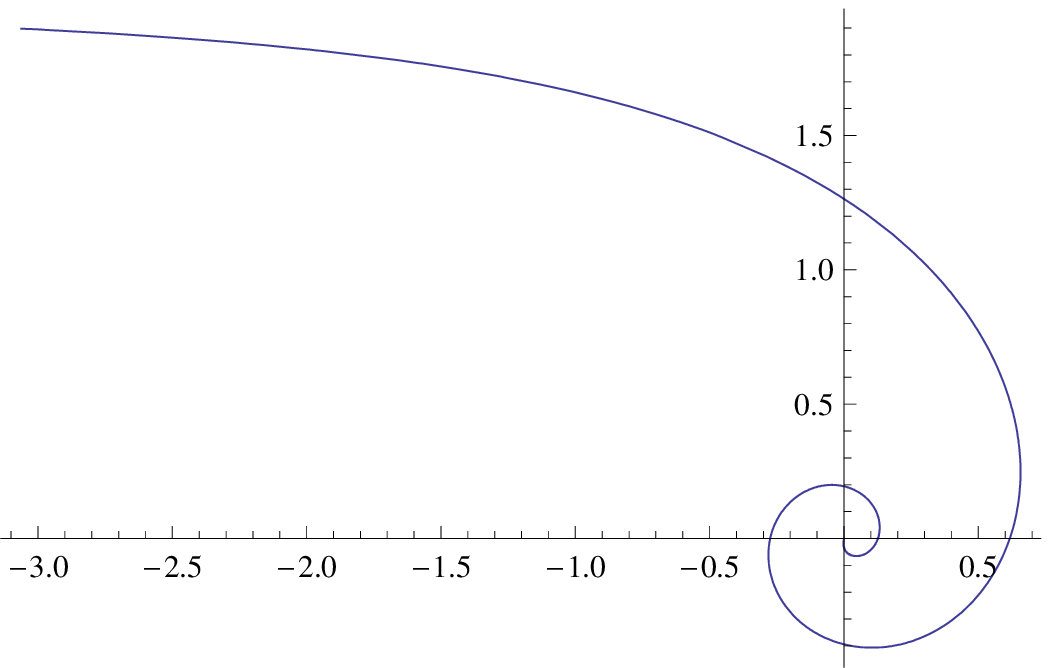}

The velocity blows up as ${1\over r^2}$ for small $r$. From a finite distance the particle is captured in a finite time.

 Other cases can be worked out similarly. The formulation of the Painlev\`e equation in terms of the Riemann-Hilbert problem and Isomonodromy\cite{Fokas,Deift}  give powerful techniques to study our solution. In particular,  there are a pair of conserved quantities that replace energy and angular momentum in this dissipative but integrable system.  Also,  we can generalize the Rutherford formula for Coulomb scattering to include radiation. We hope to return to such a detailed analysis in  a longer paper.

\section{Acknowledgements}

I thank J. Golden, S. Iyer and A. Jordan for discussions. Also thanks to H. Spohn and F. Rohrlich for comments on an earlier version of this manuscipt. This work was supported in part by the Department of Energy   under the  contract number  DE-FG02-91ER40685.

\end{document}